\documentclass{Interspeech}
\usepackage[subrefformat=parens]{subcaption}
\usepackage{multicol}
\usepackage{multirow}

\interspeechcameraready

\title{Voice Impression Control in Zero-Shot TTS}

\author[affiliation={}]{Kenichi}{Fujita}
\author[affiliation={}]{Shota}{Horiguchi}
\author[affiliation={}]{Yusuke}{Ijima}

\affiliation[nocounter]{}{NTT, Inc.}{Japan}
\email{kenichi.fujita@ntt.com}
\keywords{speech synthesis, zero-shot TTS, speaker embeddings, voice impression}

\usepackage{comment}

\begin{document}
\setlength\textfloatsep{7pt} 
\setlength\dbltextfloatsep{7pt} 
\setlength\floatsep{7pt} 
\setlength\abovecaptionskip{2pt} 
\setlength\belowcaptionskip{2pt} 
\captionsetup[subfloat]{aboveskip=2pt,belowskip=4pt} 

\maketitle

\begin{abstract}
Para-/non-linguistic information in speech is pivotal in shaping the listeners' impression. Although zero-shot text-to-speech~(TTS) has achieved high speaker fidelity, modulating subtle para-/non-linguistic information to control perceived voice characteristics, i.e., impressions, remains challenging. We have therefore developed a voice impression control method in zero-shot TTS that utilizes a low-dimensional vector to represent the intensities of various voice impression pairs (e.g., dark--bright). The results of both objective and subjective evaluations have demonstrated our method's effectiveness in impression control. Furthermore, generating this vector via a large language model enables target-impression generation from a natural language description of the desired impression, thus eliminating the need for manual optimization.
\end{abstract}

\section{Introduction}
Speech contains both linguistic and para-/non-linguistic information~\cite{cowen2019mapping}. Para-/non-linguistic information adds variations to the spoken content, shaping the listeners' impressions.
To achieve human-like expressive speech synthesis, control over various voice qualities, speaking styles, and perceived impressions is essential. While advances in zero-shot text-to-speech~(TTS) synthesis have enabled natural-sounding speech generation from a few seconds of reference speech~\cite{cooper2020zero,wang2023neural,fujita2023zeroshot}, its controllability is still challenging. Most zero-shot TTS systems struggle to control the impression of the generated utterances and can only mimic the speaker's characteristics and the speaking style of the provided reference speech. 

This challenge arises because the representations extracted from the reference speech are often entangled with factors such as pitch, rhythm, and speaking style, thereby hindering intuitive control. In response, efforts have been made to disentangle prosody, content, and acoustic details in speech representations~\cite{polyak21_interspeech,ju2024naturalspeech}. However, these methods only disentangle above basic features, thus limiting intuitive control over high-level voice impressions. Another promising approach involves leveraging emotion-related auxiliary information~\cite{9747987,pan21d_interspeech,sivaprasad21_interspeech,kang23_interspeech,INOUE202135}. Methods in this vein have demonstrated some effectiveness even in zero-shot scenarios~\cite{kang23_interspeech,cho2024emosphere}. However, emotions represent only a narrow subset of the broader characteristics that shape voice impressions, and as such, they are insufficient to control the full range of impressions. Using simple voice qualities is similarly insufficient~\cite{Frederik2025perceptual}. 

For diverse speech control, Tachibana et al.~\cite{tachibana06_interspeech} utilized voice impressions to adapt parameters in an HMM-based TTS method. However, parameter adaptation is unsuitable for zero-shot TTS. Recent text-based conditioning methods~\cite{10096285,kawamura24_interspeech,SKing_ICASSP_2024,10534832} also aim for diverse speech control via various text descriptions, but most have focused on controlling both speaker characteristics and styles, without targeting style control for specific speakers. Furthermore, text-based description is often insufficient for precise control of fine details in the generated speech.

To address this limitation, we propose a zero-shot TTS method for voice impression control. The key idea is to use a low-dimensional voice impression vector in which each dimension represents the intensity of an antonym pair describing impressions (e.g., ``dark--bright''). This vector enables intuitive and flexible voice impression control. To integrate speaker and voice impression information, we introduce a control module that removes impression-related information from the speaker representation and then reintroduces it based on the impression vector. Audio examples are available on our demo page\footnote{\url{https://ntt-hilab-gensp.github.io/is2025voiceimpression/}\label{foot:sample_page}}.

\begin{figure}[tb]
  \centering
  \includegraphics[width=\linewidth]{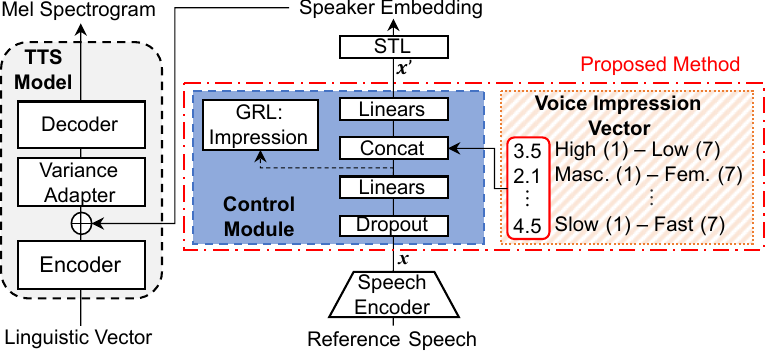}
  \caption{Overview of proposed method.}
  \label{fig:overview_propose}
\end{figure}

\section{Proposed method}
An overview of the proposed TTS method is shown in Fig.~\ref{fig:overview_propose}. We conducted the training in two steps to ensure stability. In the first step, we pre-train a zero-shot TTS model excluding the control module to obtain a high-quality TTS model. In the second step, we incorporate the proposed control module to integrate speaker and voice impression information. We train only the control module while keeping the other modules frozen. This section explains the backbone TTS model, voice impression vector, control module, and voice impression vector generation via a large language model (LLM).

\subsection{Backbone zero-shot TTS model}
\label{sec:ssl_model}

The model utilizes the target speaker's embedding extracted from reference speech to condition the TTS model. To extract the \textit{speaker embedding}, we used a speech encoder and style token layer (STL)~\cite{pmlr-v80-wang18h}. The speech encoder extracts representations using a self-supervised learning (SSL) speech model~\cite{fujita2023zeroshot} and aggregates them into a single embedding $\boldsymbol{x}$ via a weighted sum~\cite{chen22g_interspeech}, a bidirectional LSTM, and an attention mechanism~\cite{raffel2016feedforward, ando2018soft}. The STL further transforms the embedding $\boldsymbol{x}$ to ensure its stability, resulting in the \textit{speaker embedding}. Joint training of the TTS model and these modules ensures that the generated \textit{speaker embedding} is well-suited for the TTS model. 

\subsection{Voice impression vector and control module}
\label{sec:controlling_module}
Controlling voice impression is challenging because various types of speaker information---including impression information---are entangled in the embedding $\boldsymbol{x}$ extracted using a pre-trained speech encoder. To achieve flexible voice impression control, we propose two key components to 1) remove impression information coming from the embedding $\boldsymbol{x}$ and 2) reintroduce it via the impression vector. 

For the first purpose, we apply adversarial learning with a gradient reversal layer (GRL)~\cite{pmlr-v37-ganin15} to remove impression information from the embedding $\boldsymbol{x}$, following the emotional TTS studies~\cite{9747987,kang23_interspeech}.
We also apply dropout with a high-ratio (0.8 in this paper) to the embedding $\boldsymbol{x}$ for disentanglement~\cite{NEURIPS2018_03e7ef47,ICASSP2020_Ranghuveer_Rosust}.
This encourages the system to rely on the impression vector for impression-related information while using $\boldsymbol{x}$ solely to preserve target speaker characteristics.

The impression vector utilized for the second purpose is a low-dimensional vector in which each dimension quantifies the degree or intensity of a certain impression pair. We use the following ten pairs of antonyms to describe impressions covering common voice quality expressions~\cite{Mizuki_Nagano2024e24.14,kasuya1999_extraction}: A) High--Low pitched, B) Masculine--Feminine, C) Clear--Hoarse, D) Calm--Restless, E) Powerful--Weak, F) Youthful--Elderly, G) Thick--Thin, H) Tense--Relaxed, I) Dark--Bright, and J) Cold--Warm. To expand the controllable aspects of voice impressions, we added another pair, K) Slow--Fast, related to speaking rate.

\subsection{Voice impression vector generation with automatic mapping from freely described impressions using LLM}

\begin{figure}[tb]
  \centering
  \includegraphics[width=1.0\linewidth]{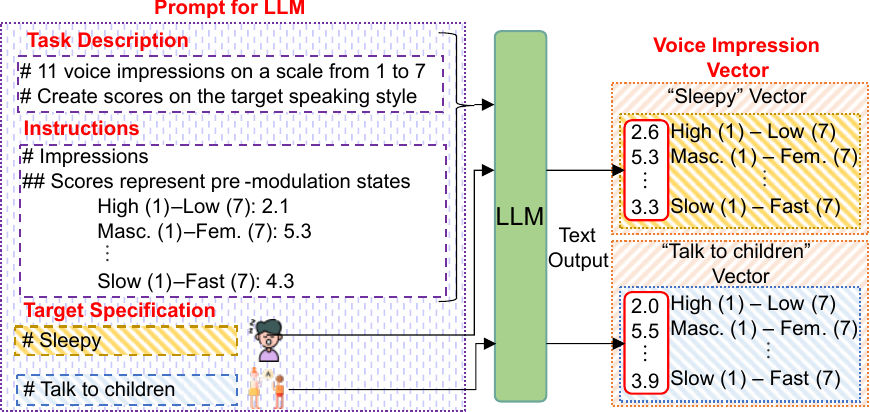}
  \caption{Voice impression vector generation via a LLM. Prompt examples are available on our demo page\footref{foot:sample_page}.}
  \label{fig:overview_LLM}
\end{figure}

The impression vector allows for fine-grained adjustments to the target voice impression. However, manual tuning to obtain speech with the desired impression is labor-intensive and requires expert knowledge, especially when the desired impression significantly differs from the source speech or when additional information (e.g., long-context details) must be considered. Inspired by LLM-based text-guided image generation~\cite{lian2023llmgrounded}, we propose leveraging an LLM to generate impression vectors automatically (Fig.~\ref{fig:overview_LLM}). The prompt comprises a task description, instructions, and the target specification. Text-based instructions, designed to improve the consistency of the LLM predictions, define each dimension and its corresponding scores for the utterance prior to modulation.

The inherent adaptability of LLMs enables vector generation using text-only prompts, effectively incorporating additional elements such as context, speaking style, and dialogue history. A previous TTS study utilized LLMs to adjust low-level speech features such as pitch, energy, and duration on the basis of context~\cite{SKing_ICASSP_2024}. Our method expands the application of LLMs to generate high-level features, i.e., voice impression vectors. Unlike prompt-based TTS methods that generate speech in a single step, our approach offers more granular control by allowing minimal manual adjustments to the impression vector after an initial global style approximation guided by the LLM. 

\begin{table*}[tb]
  \begin{minipage}{0.59\linewidth} 
    \caption{Correlation coefficients among dimensions in voice impression vector.    Values in bold indicate strong positive or negative correlation.}
    \setlength{\tabcolsep}{2pt}
    \label{table:correlation}
    \centering
    \resizebox{\linewidth}{!}{%
    \begin{tabular}{@{}c@{)\ }l*{11}{r}@{}}
    \toprule
      \multicolumn{2}{c}{}&\multicolumn{1}{c}{A} & \multicolumn{1}{c}{B}    &\multicolumn{1}{c}{C}    &\multicolumn{1}{c}{D}    &\multicolumn{1}{c}{E}    &\multicolumn{1}{c}{F}    &\multicolumn{1}{c}{G}    &\multicolumn{1}{c}{H}    &\multicolumn{1}{c}{I}    &\multicolumn{1}{c}{J}    &\multicolumn{1}{c}{K}\\ 
      \midrule
      A& High--Low pitched&1.00\\ 
      B& Masculine--Feminine& \textminus0.67 & 1.00 \\
      C& Clear--Hoarse& \textminus0.61 & 0.36 & 1.00 \\
      D& Calm--Restless& \textminus0.44 & 0.13 & 0.07 & 1.00\\
      E& Powerful--Weak& 0.49 & \textminus0.17 & \textminus0.56 & \textminus0.25 & 1.00\\
      F& Youthful--Elderly& 0.58 & \textminus0.22 & \textminus0.64 & \textminus0.37 & 0.33 & 1.00\\
      G& Thick--Thin& \textbf{\textminus0.80} & 0.67 & 0.48 & 0.35 & \textminus0.19 & \textminus0.55 & 1.00\\ 
      H& Tense--Relaxed& 0.65 & \textminus0.30 & \textbf{\textminus0.72} & \textminus0.26 & \textbf{0.79} & 0.51 & \textminus0.43 & 1.00\\ 
      I& Dark--Bright& \textbf{\textminus0.77} & 0.35 & 0.67 & 0.35 & \textminus0.67 & \textminus0.56 & 0.61 & \textbf{\textminus0.80} & 1.00\\ 
      J& Cold--Warm & \textminus0.35 & 0.11 & 0.41 & \textminus0.01 & \textminus0.23 & \textminus0.31 & 0.32 & \textminus0.39 & 0.56 & 1.00\\
      K& Slow--Fast &0.02&\textminus0.21&0.30&\textminus0.04&\textminus0.13&\textminus0.37&\textminus0.02&\textminus0.17&0.11&0.16&1.00\\
    \bottomrule
    \end{tabular}%
    }
  \end{minipage}
\end{table*}

\section{Experimental setup}

\subsection{Dataset}
\label{sec:data}
We used our in-house 1,800-hour \SI{22}{\kHz} Japanese speech database comprising 20,270 speakers. It was split into two parts: 1,588,847 utterances from 20,222 speakers for training and 7,278 utterances from 48 speakers for validation. All utterances were labeled with an 11-dimensional voice impression vector (see details in Sect.~\ref{sec:data_voce_impression}).
\subsection{Collection of voice impression vectors}
\label{sec:data_voce_impression}

The scores for each dimension A) through J) of the voice impression vectors were obtained via subjective evaluation. Given the impracticality of subjective evaluations for all utterances due to time and cost constraints, we adopted automatic labeling using a three-step process: crowdsourcing subjective scores, training an impression vector estimator, and estimating vectors for all utterances. The K) Slow--Fast scores were calculated by standardizing speech rates~(in moras per second) across the entire dataset. The data preparation process is detailed below. 

We began by collecting subjective evaluations from 1,154 speakers in the dataset. Each speaker's single utterance was rated on ten impression pairs, namely A) through J). Although only one utterance was evaluated per speaker, we can assume that each speaker's voice impression remained constant across other unevaluated utterances. This assumption is supported by the design of the TTS dataset, in which the speakers were instructed to maintain a consistent speaking style. Participants rated each utterance on each impression pair using a 7-point Likert scale (e.g., 1: dark to 7: bright). Each utterance was evaluated by at least ten participants, and the average score for each pair was utilized as its corresponding impression vector. 

To automatically label the remaining data, we trained an impression vector estimator using utterances and their corresponding impression vectors obtained from the subjective evaluation. The estimator architecture comprised an SSL model followed by a bidirectional LSTM and an attention mechanism. For the SSL model, we used the publicly available HuBERT \textsc{Base}~\cite{hsu2021hubert} trained on ReazonSpeech~\cite{fujimotoreazonspeech}\footnote{\url{https://huggingface.co/rinna/japanese-hubert-base}\label{foot:hubert_reazon}}. The model was trained to minimize the mean squared error (MSE). The training data comprised 146,809 utterances from 1,067 speakers and 6,151 validation data from 44 speakers. All these utterances respectively belong to the training and validation data described in Sect.~\ref{sec:data}. The estimator was trained for 30 epochs, and the epoch with the lowest validation MSE was selected. For testing, we evaluated the model using 5,719 utterances from 43 speakers not included in the training or validation data. The root MSE was 0.338, indicating high estimation accuracy. After training, we applied the estimator to utterances that had not been subjectively evaluated and obtained their impression vectors. Note that the impression pairs are not independent. Table~\ref{table:correlation} lists the correlations among the dimensions, calculated using impression vectors obtained from subjective evaluations instead of the estimated data.

\begin{figure}[tb]
  \centering
  \includegraphics[width=1.0\linewidth]{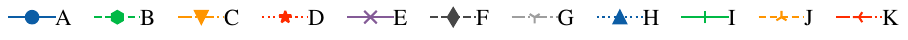}\\  
  \subfloat[\textit{Speaker 1} (female)]{%
    \includegraphics[width=0.48\linewidth]{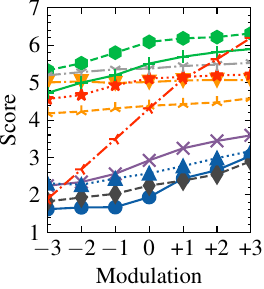}%
    \label{fig:obeval_imp_male}%
    }
  \hfill
  \subfloat[\textit{Speaker 2} (male)]{%
    \includegraphics[width=0.48\linewidth]{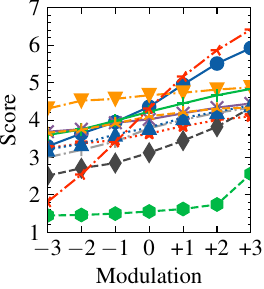}%
    \label{fig:obeval_imp_female}%
  }
  \caption{Objective evaluation results for single-dimension modulation of the voice impression vector.}
  \label{fig:ob_eval_emp_1d}
\end{figure}

\subsection{Training conditions}
The TTS model was FastSpeech2~\cite{ren2020fastspeech}, as implemented in a previous study~\cite{chien2021investigating}, which features four and six feed-forward Transformer blocks in the encoder and the decoder, respectively. The inputs were 303-dimensional linguistic vectors, while the target was an 80-dimensional log-mel spectrogram with a frame shift of \SI{10.0}{ms}. We used HuBERT \textsc{Base} trained on ReazonSpeech\footref{foot:hubert_reazon} as the SSL model, keeping its parameters frozen. HuBERT processed raw audio input sampled at \SI{16}{kHz} into a 768-dimensional vector sequence, which is further converted into fixed-length 384-dimensional \textit{speaker embedding}.  In the control module, the embedding $\boldsymbol{x}$ (384 dimensions) and voice impression vectors (11 dimensions) were projected to 32 dimensions. Dropout was applied with a rate of 0.8. These hyper-parameters were determined empirically. For waveform generation, we employed HiFi-GAN (V1)~\cite{NEURIPS2020_c5d73680}. 

To ensure stable training, we initially pre-trained the backbone TTS model for 200k steps without the control module. We used the Adam optimizer~\cite{kingma-Adam} with the Noam scheduler~\cite{NIPS2017_3f5ee243}. Subsequently, we further trained the backbone TTS model using a GAN~\cite{kanagawa19_ssw,peng2018variational} for an additional 200k steps at a fixed learning rate of 0.001 to improve quality. Finally, we inserted the control module and trained it with the same training data for an additional 50k steps. During this stage, we employed the same GAN with its associated loss function for adversarial training, and all modules except the control module were frozen. 

\section{Results}
\subsection{Objective evaluation}
We conducted three objective evaluations to verify impression controllability and its effect on speaker similarity. We selected two speakers---\textit{Speaker 1}~(female) and \textit{Speaker 2}~(male)---not present in training or validation data, and chose one utterance from each as the reference speech. This implies that the generation is in a zero-shot condition. We generated 50 utterances per condition in each of the following experiments. To evaluate the degree of change in impression, we repurposed the impression estimator, originally prepared for data generation. 

First, we modulated one dimension of the impression vector corresponding to the reference speech. The modulation ranged from \textminus3 to +3 relative to its original value. The proposed model takes a voice impression vector itself as input, but from the perspective of impression controllability, we report the results based on the modulation levels of the vector. Figure~\ref{fig:ob_eval_emp_1d} illustrates the changes in the impression of the modulated dimensions. The consistent upward trend across all dimensions for both \textit{Speaker~1} and \textit{Speaker~2} indicates that changes in the impression vector correspond to changes in the impression. 

\begin{figure}[t]
  \begin{flushright}
  \vspace{-49mm}%
  \begin{minipage}{0.82\linewidth}
  \centering
  \includegraphics[width=0.8\linewidth]{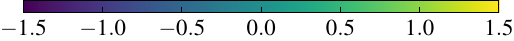}\\
  \subfloat[Dim. E (Powerful--Weak)]{%
      \makebox[0.49\linewidth][c]{%
      \includegraphics[width=0.48\linewidth]{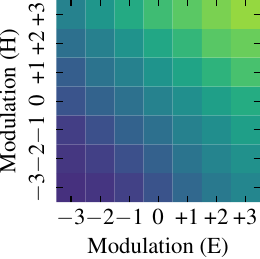}%
      }%
  }
  \hfill
  \subfloat[Dim. H (Tense--Relaxed)]{%
      \includegraphics[width=0.48\linewidth]{./pictures/heatmap_2d_E.pdf}%
  }
  \caption{Change in scores when two dimensions~(E and H) are simultaneously modulated.}
  \label{fig:ob_eval_emp_2d}
  \end{minipage}
\end{flushright}
\end{figure}

Second, we examined simultaneous two-dimensional modulation because, in many cases, we must modulate more than one dimension at the same time to achieve the desired impression using manual or LLM modulation. We selected dimensions E and H because of their high correlation (Table~\ref{table:correlation}). The modulation level ranged from \textminus3 to +3. We then analyzed the changes in the corresponding impression vector dimensions (E and H) estimated from the generated speech. Figure~\ref{fig:ob_eval_emp_2d} presents the results for \textit{Speaker 1}, where it is clear that even with the simultaneous modulation, the generated speech consistently reflected the intended changes. Note that the same tendency was observed for \textit{Speaker 2} and other dimensions. These experiments demonstrate that our proposed method successfully alters the impressions corresponding to the modulated dimensions.

\begin{figure}[tb]
  \centering 
  \subfloat[Results for \textit{Speaker 1 (female)}]{%
  \includegraphics[width=0.95\linewidth]{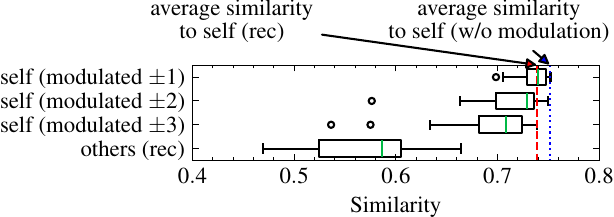}
  }\hfill
  \subfloat[Results for \textit{Speaker 2 (male)}]{%
  \includegraphics[width=0.95\linewidth]{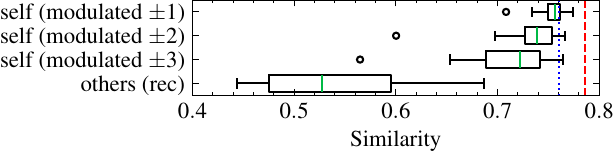}
  }
  \caption{Speaker similarity to target speaker's reference speech evaluated with Resemblyzer. ``self'' denotes generated utterances with and without modulation, with levels of $\pm1$, $\pm2$, and $\pm3$. ``self/others (rec)'' denotes different recorded speech of the same speaker and the recorded speech of different speakers of the same gender, respectively.}
  \label{fig:obj_similarity}
\end{figure}

Finally, we used Resemblyzer\footnote{\url{https://github.com/resemble-ai/Resemblyzer}} to evaluate the impact of modulation on speaker identity. In Fig.~\ref{fig:obj_similarity}, we analyzed how the cosine similarity between the generated utterance and the reference speech transitioned as the modulation level changed from 0 to $\pm1$, $\pm2$, and $\pm3$. For comparison, we also show the average cosine similarity to different recorded speech of the same speaker, as well as the distribution of cosine similarities to that of different speakers of the same gender (20 female and 21 male speakers, respectively). As shown, greater modulation reduced similarity; however, modulated utterances were still more similar than those from different speakers. This suggests that the proposed method achieves noticeable changes in voice impression while minimizing deterioration in speaker similarity. Some outliers are due to large changes in pitch, e.g., resulting from modulation in the B)~Masculine--Feminine dimension.

\begin{table}[tb]
    \caption{Results of subjective evaluation on voice impression scores with 95\% confidence interval.}
    \setlength{\tabcolsep}{4pt}
    \label{table:sub_impression}
    \centering
    \resizebox{\linewidth}{!}{%
    \begin{tabular}{@{}l@{\hspace{2pt}}c*{4}{c}@{}}
        \toprule
        &&\multicolumn{4}{c}{Modulation level}\\\cmidrule(l{\tabcolsep}){3-6}
       Speaker&Dim.& \textminus3 & \textminus1.5 & +1.5 &+3 \\
        \midrule
        \multirow{4}{*}{\textit{Speaker 1}}&  B & $3.39\pm0.19$     &$3.89\pm0.18$  &$4.14\pm0.20$ &$4.33\pm0.24$\\
        &   E  & $3.44\pm0.14$     & $3.61\pm0.11$  &$4.30\pm0.12$&$4.68\pm0.13$\\
        &   F &  $3.35\pm0.17$    & $3.61\pm0.16$  & $5.03\pm0.14$&$5.70\pm0.15$\\
        &   I  & $3.03\pm0.17$     &$3.48\pm0.14$   & $4.59\pm0.14$&$4.85\pm0.19$\\\midrule
        \multirow{4}{*}{\textit{Speaker 2}}&  B &  $3.45\pm0.21$&$3.71\pm0.20$&$4.46\pm0.18$ &$4.71\pm0.17$\\
        &   E  & $3.37\pm0.14$& $3.68\pm0.10$  &$4.38\pm0.12$&$4.71\pm0.16$\\
        &   F & $3.86\pm0.15$& $3.97\pm0.12$  &$4.08\pm0.12$ &$4.54\pm0.16$\\
        &   I  & $3.20\pm0.15$& $3.56\pm0.12$ &$4.42\pm0.12$ &$4.66\pm0.14$\\
        \bottomrule
    \end{tabular}%
    } 
\end{table}

\subsection{Subjective evaluation}
We conducted subjective evaluations of the controllability and naturalness of the proposed method. Referring to Table~\ref{table:correlation}, we selected four weakly correlated dimensions: B)~Masculine--Feminine, E) Powerful--Weak, F) Youthful--Elderly, and I)~Dark--Bright. We then modulated each impression vector dimension using the values \textminus3.0, \textminus1.5, +1.5, and +3.0. For each modulation level, we synthesized 12 sentences for each speaker, \textit{Speaker 1} and \textit{Speaker 2}. The experiments for controllability and naturalness were conducted on a crowdsourcing platform with 478 and 178 participants, respectively, and each utterance was evaluated at least ten times.

To evaluate controllability, each participant was instructed to listen to pairs of samples---each consisting of one with voice impression modulation and one without---and rate the impression difference using a 7-point Likert scale (e.g., 1:~dark to 7:~bright). The order of the two samples was randomized. Table~\ref{table:sub_impression} lists the results. The changes in voice impression align with the respective modulations, with larger modulation widths resulting in greater impression changes.

The naturalness was evaluated using a mean opinion score~(MOS) on a five-point scale (5: very natural to 1: very unnatural). Naturalness was assessed to detect quality degradation introduced by the impression modulation. The results in Table~\ref{table:MOS} show some degradation in naturalness but most speech samples with low naturalness scores exhibited outlier expressions, such as ``feminine male,'' ``weak,'' and ``strongly dark/bright'' utterances. Compared to neutral utterances, their expressions could convey less naturalness.

\begin{table}[tb]
    \caption{Results of subjective evaluation on naturalness scores with 95\% confidence interval. Bold numbers indicate $p<.05$ in the difference to No modulation (0).}
    \setlength{\tabcolsep}{4pt}
    \label{table:MOS}
    \centering
    \resizebox{\linewidth}{!}{%
    \begin{tabular}{@{}l@{\hspace{2pt}}c*{5}{c}@{}}
        \toprule
        &&\multicolumn{5}{c}{Modulation level}\\\cmidrule(l{\tabcolsep}){3-7}
       Speaker&Dim.& \textminus3 & \textminus1.5 & 0 &+1.5 &+3 \\
        \midrule
        \multirow{4}{*}{\textit{Speaker 1}}%
        &   B  & $3.55\pm0.22$         &$3.74\pm0.18$   & $3.72\pm0.18$    &$3.84\pm0.19$ &$3.66\pm0.19$\\
        &   E  & $3.62\pm0.20$     & $3.72\pm0.19$ &$3.72\pm0.18$  &$\bf 3.36\pm0.22$  &$\bf 3.18\pm0.22$\\
        &   F  &  $\bf 3.20\pm0.19$    & $\bf 3.34\pm0.20$  & $3.72\pm0.18$    &$3.72\pm0.19$  &$3.45\pm0.22$\\
        &   I  & $\bf 3.02\pm0.25$     &$\bf 3.18\pm0.21$   & $3.72\pm0.18$& $3.55\pm0.18$ &$\bf 2.71\pm0.24$\\\midrule
        \multirow{4}{*}{\textit{Speaker 2}}&  B &  $3.51\pm0.18$&$3.56\pm0.16$&$3.61\pm0.14$&$3.46\pm0.18$ &$\bf 3.07\pm0.20$\\
        &   E  & $\bf 3.35\pm0.19$& $3.64\pm0.17$  &$3.61\pm0.14$&$\bf 3.21\pm0.18$&$\bf 2.88\pm0.20$\\
        &   F & $3.46\pm0.20$& $3.60\pm0.16$  &$3.61\pm0.14$&$3.43\pm0.18$ &$\bf 3.23\pm0.17$\\
        &   I  & $\bf 3.18\pm0.19$& $\bf 3.33\pm0.17$ &$3.61\pm0.14$&$3.38\pm0.18$ &$\bf 2.78\pm0.22$\\
        \bottomrule
    \end{tabular}%
    }
\end{table}

\begin{table}[tb]
  \caption{Preference scores (\%) of AB test between LLM-based generation (left) vs. no modulation (right). Bold numbers indicate the significantly preferred method with $p<.05$.}
  \label{table:AB}
  \centering
  \resizebox{\linewidth}{!}{%
  \begin{tabular}{@{}lS[table-format=2.1]@{~vs.~}S[table-format=2.1]S[table-format=2.1]@{~vs.~}S[table-format=2.1]S[table-format=2.1]@{~vs.~}S[table-format=2.1]@{}}
    \toprule
    Target impression& \multicolumn{2}{c}{\textit{Speaker 1}}&\multicolumn{2}{c}{\textit{Speaker 2}} & \multicolumn{2}{c@{}}{All} \\
    \midrule
    \textit{Sleepy} & \textbf{98.5} &  1.5& \textbf{91.0} & 9.0& \textbf{94.6} &  5.4\\
    \textit{Urgent, attention grabbing} & \textbf{67.4} & 32.6& \textbf{81.9} & 18.1 & \textbf{74.6} & 25.4\\
    \bottomrule
  \end{tabular}%
  }
\end{table}

\subsection{Evaluation of voice impression vector generation using an LLM}
We conducted a subjective evaluation of voice impression generation using an LLM. We selected one utterance each from \textit{Speaker 1} and \textit{Speaker 2}, and generated them to convey the impressions of ``sleepy'' and ``urgent, attention-grabbing,'' respectively. These impressions were chosen because their complexity exceeds the emotion-based methods or simple dimensional adjustments in this study. Although we utilized simple expressions for simplicity, more detailed contexts and scenarios could be specified. We used ChatGPT-4o~\cite{gpt4o} for the LLM.

Subjective evaluations were conducted via crowdsourcing with 436 participants. In the test, participants listened to pairs of samples: one generated using LLM and one without. For each pair, they selected one sample that better matched the target impression. The experimental results, summarized in Table~\ref{table:AB}, show a significant preference for the sample with the target impression generated using LLM. These results demonstrate that leveraging an LLM enables us to achieve the desired voice expressions without direct vector modulation. Audio examples and generated impression vectors from the LLM are available on our demo page\footref{foot:sample_page}.

\section{Conclusion}
We introduced a voice impression control method in zero-shot TTS utilizing a voice impression vector. Objective and subjective evaluations demonstrated that the proposed method controls voice impression while preserving speaker similarity to the reference speech. Moreover, our findings confirmed the feasibility of controlling voice impressions with an LLM via the proposed vector. Future work will involve voice impression constraints on neural audio codecs---similar to the approach used in NaturalSpeech3~\cite{ju2024naturalspeech}, which emphasizes prosody and acoustic details---and evaluating the effectiveness.

\clearpage
\bibliographystyle{IEEEtran}
\bibliography{refs}

\end{document}